\documentclass[numbered]{trbunofficial}

\usepackage{xcolor}
\usepackage{graphicx}
\usepackage{amsmath,amsfonts,amssymb,amsthm,epsfig,epstopdf,url,array}
\usepackage{amsfonts}
\usepackage{float}
\usepackage{subfig}
\theoremstyle{plain}

\theoremstyle{remark}

\AuthorHeaders{Park, Xue, and Ukkusuri}
\title{Finding critical transitions of the post-disaster recovery using the sensitivity analysis of agent-based models}

\author{
  \textbf{Sangung Park}\\
  Lyles School of Civil Engineering\\
  Purdue University, West Lafayette, IN, 47907\\
  Email: park1202@purdue.edu\\
  \hfill\break
  \textbf{Jiawei Xue}\\
  Lyles School of Civil Engineering\\
  Purdue University, West Lafayette, IN, 47907\\
  Email: xue120@purdue.edu\\
  \hfill\break 
  \textbf{Satish V. Ukkusuri, Ph.D.}\\
  Lyles School of Civil Engineering\\
  Purdue University, West Lafayette, IN, 47907\\
  Email: sukkusur@purdue.edu\\
  (Corresponding Author)
}

\TotalWords{5987}
\begin{document}
\maketitle

\section{Abstract}
Frequent and intensive disasters make the repeated and uncertain post-disaster recovery process. Despite the importance of the successful recovery process, previous simulation studies on the post-disaster recovery process did not explore the sufficient number of household return decision model types, population sizes, and the corresponding critical transition conditions of the system. This paper simulates the recovery process in the agent-based model with multilayer networks to reveal the impact of household return decision model types and population sizes in a toy network. After that, this paper applies the agent-based model to the five selected counties affected by Hurricane Harvey in 2017 to check the urban-rural recovery differences by types of household return decision models. The agent-based model yields three conclusions. First, the threshold model can successfully substitute the binary logit model. Second, high thresholds and less than 1,000 populations perturb the recovery process, yielding critical transitions during the recovery process. Third, this study checks the urban-rural recovery value differences by different decision model types. This study highlights the importance of the threshold models and population sizes to check the critical transitions and urban-rural differences in the recovery process.
\\ 

\hfill\break
\noindent\textit{Keywords}: Post-disaster recovery, Agent-based modeling, Multilayer network, Critical transitions, Threshold models, Hurricane Harvey
\newpage

\section{Introduction}
A post-disaster recovery (PDR) process is generally defined as a process of overall human environment recovery activities after disasters, including physical, social, and environmental recovery events within affected communities~\cite{rouhanizadeh2020exploratory}. 
The PDR process is essential for mitigating the consequent risks of natural disasters. A 40\% increase in the estimated natural disasters from 2015 to 2030 necessitates the fast PDR process to provide sufficient time for consequent disaster responses and preparedness~\cite{undrr2022global,ipcc2023}. Otherwise, consequent disasters can evolve into compound disasters whose impacts are overlapped and enlarged throughout the disaster periods~\cite{liu2014compound}, increasing the physical and social vulnerability~\cite{mcgreevy2023second}. 

Previous studies on the PDR process have explored the influential factors for the fast process, such as housing damage~\cite{lee2022patterns,griego2020social}, the functionality of physical~\cite{2019census}, social~\cite{aldrich2012building}, or coupled socio-physical infrastructures~\cite{yabe2021resilience}, neighbor activities~\cite{yabe2022toward}, the past experiences on the natural disasters ~\cite{rouhanizadeh2020exploratory}, recovery information from governments~\cite{rouhanizadeh2020exploratory}, recovery assistance~\cite{sadri2018role}, sociodemographic~\cite{sadri2018role}, and socioeconomic attributes across regions~\cite{hasan2011behavioral}. These interdependent interactions among the influential factors make the PDR process complex~\cite{yu2022digital,fan2021disaster}.

There are two types of simulations and techniques to analyze the complex return decision after natural hazards. Specifically, past work has used system dynamics (SDs) with a high abstraction of the systems and homogeneity of the elements, meaning that all elements follow the suggested dynamics~\cite{yabe2021resilience}. On the other hand, agent-based models (ABMs) have been used to explain a low abstraction of the systems and heterogeneity of the agents~\cite{moradi2020recovus,hajhashemi2019using,ghaffarian2021agent}. The SD on the PDR process utilized the nonlinear functions, such as the logistic functions~\cite{yabe2021resilience,sutley2018interdisciplinary} or Hill type functions~\cite{yabe2021resilience} to model the nonlinear complex interactions in the systems. However, previous studies lost the opportunity to study the heterogeneous impacts of the elements. This is because the SD model has the homogeneity assumption that each element utilizes the same relationships among the other factors. 

The homogeneity assumption of the SD prompts a wide usage of ABMs for the PDR process since ABMs are free from the homogeneity~\cite{esmalian2022multi}. The ABM is composed of agents, environments, and actions. Agents heterogeneously interact with other elements in the environment to simulate the complex system. On the PDR process, ABMs usually adopt the logistic function as an agent's decision rule to express nonlinear and heterogeneous relationships among agents and the other elements~\cite{hajhashemi2019using} based on the utility function-based decision rules~\cite{ben1985discrete}. 

Nonetheless, the ABM's utility function-based logistic function assumes commensurability of attributes which is not applicable to the PDR process. The commensurability of attributes means that they are interchangeable, so they are reducible to one attractiveness measure, called by a utility~\cite{ben1985discrete,svenson1979process}. For instance, the lack of sufficient recovery status of the social infrastructure can be interchangeable with the perfect recovery status of the physical infrastructure or the number of neighbors. However, we experienced counterexamples of the commensurability of attributes, such that the attributes around the PDR process are not always interchangeable. For example, the high recovery status of the physical infrastructure cannot complement the low recovery status of the social infrastructure. In other words, although the physical infrastructure is almost fully recovered, the slow recovery of the social infrastructure prevents households from returning home. This phenomenon has been observed in the PDR process after Hurricane Katrina~\cite{groen2010going,falk2006hurricane}, Sandy~\cite{bryner2017washed}, and Maria~\cite{roman2019satellite}. 

The counterexamples of the commensurability of attributes in the PDR process trigger the introduction of three consequent concepts: (1) a threshold model, (2) a multilayer network (MN), and (3) a critical transition (CT). 
The threshold model uses several threshold values to determine the discontinuous states of the agent~\cite{macy2020threshold,suding2009threshold}. Since the threshold model fits the nonlinear relationships and inherently blocks the commensurability of attributes~\cite{suding2009threshold}, the threshold model can successfully explain the return decision of the agents in the PDR process. 

The next concept, the MN, is necessary to complement the threshold model in the PDR process. Although the threshold model is not interchangeable among the attributes on the agent's return decision model, the other studies observed the macroscopic connections between the physical and social infrastructures~\cite{yabe2021resilience} outside of the household's return decision. It motivates researchers to use the MN to identify the interactions among the elements of the system exterior to the return decision model. The MN is a set of network layers containing distinguishable types of interactions between and within the layers. The MN is mandatory for the PDR process to clearly state different kinds of interactions inside and outside of the return decisions in the PDR process. Additionally, the MN has been widely used in simulating complex systems~\cite{boccaletti2014structure}.


Two concepts of the threshold model and the MN yield the CTs. The CTs denote abrupt changes from one state to the other state by the small perturbation of the elements~\cite{scheffer2020critical,scheffer2009early,george2021early}, and one of the emergent patterns coming from multi-stability of the complex system~\cite{george2021early,kuehn2011mathematical}. The CTs have been widely used in physics~\cite{mayergoyz2003mathematical} and ecological systems~\cite{scheffer2020critical,scheffer2001catastrophic}. CTs are nowadays emerging in the disaster resilience area~\cite{yabe2022toward} since the disaster resilience area follows the concept of complex systems. 
The threshold model and the MN are mainly used to simulate the complex system~\cite{george2021early,scheffer2009early}, so researchers should gird for the emergence of the CTs if they consider the complex systems. 
Previous studies have analyzed the CTs to reveal the condition and mathematical expressions of the emergence, and they tried to observe the early warnings of the CTs to prevent the emergence of the CTs~\cite{kuehn2011mathematical,george2021early}.
We had already observed the CTs from the recovery to the displacement phase in the PDR process when we mentioned the unusual displacement of the households in three different disaster cases~\cite{groen2010going,falk2006hurricane,bryner2017washed,roman2019satellite}. 

Throughout three consequent concepts, this paper asserts that the PDR process necessitates the threshold-based return decision model to block the commensurability of attributes. Also, analyzing the MN and the CTs is necessary to support the threshold model keeping the underlying mechanisms of the external interactions and to understand the unusual displacement of households after the disaster. Therefore, this paper introduces the agent-based models in the multilayer network (ABM-MN) to represent the threshold models of the agents in the PDR process. The ABM-MN can embrace the CTs of the PDR process since the ABM-MN is flexible to the homogeneity of the agents and emergent patterns in the complex systems.

There is room for exploring the effect of population size even though ABMs are frequently used to simulate the complex system. The population size of the ABMs affects the agent interactions among agents of ABMs, invoking several experiments with other influential factors~\cite{macy2020threshold,granovetter1978threshold}. However, little attention has been paid to the simultaneous impact of the population size and agent's return decision models in ABM-MN on the PDR process due to the limited data sources. In other words, the next PDR process does not repeat the same effects of the previous PDR process. Survey data and mobile phone location data are good foundations to represent the agent's decision models in the ABM, but these two datasets provide a limited sample of the population. Some researchers and policymakers questioned the representation of the ABMs driven by the survey data or the mobile phone location data due to the limited amount of datasets~\cite{felbermair2020generating}. If we explore the population size effect by different return decision model types, we can improve the robustness of the ABMs and check whether the ABM results represent the real world.

In summary, this study raises two research questions: 
\begin{itemize}
    \item Do the types of agent's return decision models and the agent population size affect the macroscopic patterns in the ABM-MN to represent the PDR process?
    \item Which condition brings about the CTs in the overall PDR process?
\end{itemize}


The main goal of this study is to quantify the macroscopic impact of the selected types of the agent's decision rules and population sizes in the ABM-MN to check the conditions of the CTs in the PDR process. The ABM-MN has three main building blocks, (1) the return decision rule of the human agents, (2) the MN, and (3) the SD model for the background interactions. First, this study introduces and compares various threshold models, such as a universally homogeneous threshold, a universally heterogeneous threshold, an individually heterogeneous threshold, and a universally time-varying threshold. Second, this paper applies the MN to describe the independent interactions inside the return decision models to overcome the commensurability of the attributes. Third, this study selects the SD model to explain the interactions exterior to the agents' decision rules. Note that the MN and the SD model use the results of the previous studies~\cite{xue2023supporting}.

With three building blocks of the ABM-MN, this paper simulates the PDR process in the toy network and checks the impacts of the threshold models and the population size. Previous study has raised the issue of insufficient population size of ABM~\cite{srikrishnan2021small}. Consequently, this paper explains the conditions of the CTs and tracks the cause of the CTs by simulating counterfactual scenarios with extreme conditions for the PDR process. Next, we apply the ABM-MN to five selected counties in Texas to simulate the PDR process after Hurricane Harvey, 2017. We utilize three types of data sources to estimate parameters of the ABM-MN, (1) the mobile phone location data, (2) visits to the point-of-interests (POIs), and (3) the survey data coming from the previous study~\cite{xue2023supporting}. 

\begin{figure}[h]
    \centering
    \includegraphics[scale=0.5]{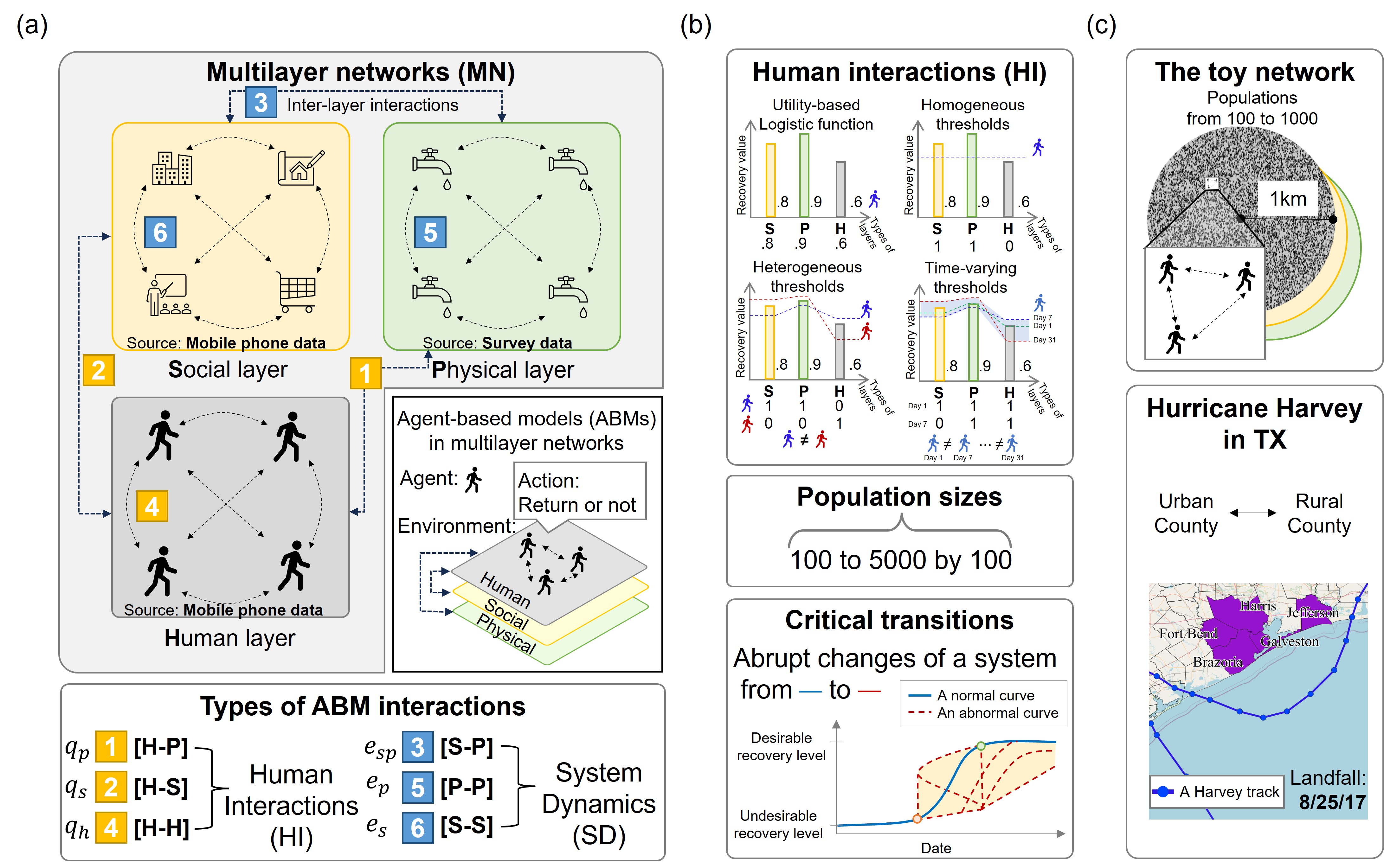}
    \caption{Schematic overview of this study. \textbf{(a)} Schematic illustration of a multilayer network and multi-agent systems (MAS). \textbf{(b)} Types of MAS interactions and focuses of this study. \textbf{(c)} Spatial and temporal scales of this study. A toy network consisting of MNs (Top). A trajectory of Hurricane Harvey in Texas, 2017 (Bottom).}
    \label{fig1}
\end{figure}

Figure \ref{fig1}a explains ABM-MNs with details on the MNs and their interactions. Figure \ref{fig1}b illustrates three main points of this study: (1) four different types of human interactions, (2) population sizes, and (3) identifying critical transitions. Figure \ref{fig1}c describes two study areas, a circle area with a 1-kilometer radius and five counties affected by Hurricane Harvey, Texas, in August 2017. 

This study highlights the importance of the threshold models and the data-driven condition of the CTs of the PDR process. First, this paper is the first study to introduce the threshold models in the PDR process based on the shortcomings of previous return decision models. 
On the other hand, this study can notify policymakers of the caution of CTs in the PDR process. Policymakers can consider the condition of CTs and make recovery efforts to close the urban-rural gaps to block the CTs of the PDR process in rural areas. 

The remaining sections of this paper are presented as follows. Section 2 describes the datasets, and Section 3 describes the overall simulation framework of ABM-MN. Section 4 explains a counterfactual scenario setup for analyzing the impact of the types of return decision rules and population sizes in a toy network. This study also discovers the condition of CTs in counterfactual scenarios. Section 5 explains the simulation results of different real-world scenarios by the model types and discusses the potential implications of CTs. Section 6 summarizes the outcomes, potential applications, and limitations of this study. Section 7 concludes this study.

\section{Data descriptions}
\subsection{Hurricane Harvey and the five counties}
Hurricane Harvey is selected as a target disaster to simulate the PDR process. Hurricane Harvey, the second-costliest U.S. hurricane until 2023, made its first landfall in Texas of the U.S. on August 25, 2017, and dissipated on September 2. Hurricane Harvey caused 780,000 refugees, 42,000 people temporarily living in 270 shelters, and 2.2 billion dollars of housing damages, as estimated by the Federal Emergency Management Agency~\cite{hapfema}. 

Five Texas coastal counties (Fort Bend, Harris, Jefferson, Brazoria, and Galveston counties) are selected as the case study areas of the ABM-MN to understand the PDR patterns. The counties are sufficient to analyze the hurricane impacts near Houston to check the urban-rural differences of the PDR process. The bottom of Figure \ref{fig1}c represents the selected five coastal counties, and Table \ref{table_demographics} shows the demographics and the number of data by the selected counties. Population and population density rows of Table \ref{table_demographics} come from the 2020 US Census database. The number of estimated homes and POIs of Table \ref{table_demographics} comes from the mobile phone location and POI data. A row with total damage on housing comes from estimation data of Hurricane Harvey for housing assistance by the Federal Emergency Management Agency database.

\begin{table*}[h]
\centering
\footnotesize
\caption{Sociodemographics and hurricane damages by the selected counties.}
\begin{tabular}{c|ccccc}
\hline
Descriptions & Harris & Fort Bend & Brazoria & Galveston & Jefferson \\
\hline
Population              & 4,731,145 & 822,779   & 372,031  & 350,682   & 256,526   \\
Population density (people per 1km radius circle) & 3,362.00  & 1,158.15  & 331.02   & 1,121.52  & 354.92    \\
The number of estimated homes          & 35,497    & 8,947     & 1,616    & 1,814     & 1,704     \\
The number of POIs                     & 66,995    & 8,947     & 4,834    & 5,330     & 4,407     \\
The percent of valid users                 & 0.75\%    & 1.09\%    & 0.43\%   & 0.52\%    & 0.66\%    \\
Total damage on housing (millions)   & 1,040.76     & 115.48     & 104.94    & 151.28   & 220.00   \\
\hline
\end{tabular}
\label{table_demographics} 
\end{table*}

\subsection{Mobile phone location data}
This study utilizes three types of data to estimate the elements of the PDR process, (1) mobile phone location data, (2) POI data, and (3) survey data. This paper will briefly discuss the usage of the data in the ABM-MN—see~\cite {xue2023supporting} for more details.

This paper utilizes Texas mobile phone location data from Quadrant in 2017. The mobile phone location data contain the detailed geographical trajectories of mobile phone users. This study extracts the return decisions of mobile phone users from the mobile phone location data to represent the populations' return decisions. To do this, this study estimated the home location of each user based on the trajectories of mobile phone users.

\subsection{POI data}
This paper exploits the POI data collected by Safegraph\footnote{ https://www.safegraph.com/}.
The POI data contain the number of visits to POIs and corresponding attributes such as open hours, brand, and POI types by North American Industry Classification System (NAICS) code. Throughout the number of visits to POIs, this paper estimates the social layer's recovery states. 

\subsection{Post-hurricane survey data}
This paper collects the post-hurricane survey data for Hurricane Harvey, 2017, to estimate household interactions. SD models, physical infrastructure recovery status (water infrastructure data), and agent's return decision models are estimated from the survey data. The purpose, template, and description of the questionnaires are described in the previous study~\cite{xue2023supporting}.

\subsection{The toy network}
This study executes the scenarios on the toy network at first. The toy network allows us to represent the standard simulation irrespective of the geolocation of POIs and the different number of neighbors. The toy network is also necessary for analyzing the conditions of the CTs, which will be described in Section \ref{sensitivity}. 

This study selects a circle area with a 1-kilometer radius of Galveston County since Galveston County contains the median of the population density of the five selected counties. This study randomly selects the number of mobile users' values from Galveston County. Locations are randomly disseminated in a circle with a 1-kilometer radius. We extract all parameters coming from the result of Galveston County~\cite{xue2023supporting} since Galveston County has the attributes of the urban and rural areas. The top of Figure \ref{fig1}c represents the toy network of three layers. The number of POIs sets to be 17, converted from the number of POIs in Galveston County.

\section{An ABM-MN simulation framework}
This study utilizes the ABM-MN simulation for the PDR process designed by~\cite{xue2023supporting}. This study uses the notations to precisely describe the interactions by Table \ref{table_notations}.

\begin{table*}[h] 
\centering
\footnotesize
\caption{The nomenclature of this study.}
\begin{tabular}{c|l}
\hline
Notations           & Descriptions                                                                                                  \\
\hline
$|\cdot|$                  & The cardinality of a set or a graph                                                                                      \\
$\bar{\cdot}$ & An average of $\cdot$ \\
$h$& an indicator of a home layer. \\
$s$ & an indicator of a social layer. \\
$p$ & an indicator of a physical layer. \\
$i, k$ & an indicator of layers. $\forall i,k\in\{h,s,p\}$ \\
$j$ & an indicator of agents in $|\cdot|$. $\forall j\in\{1,2\mathellipsis,|\cdot|\}$ \\
$l$ & an indicator of regions. $\forall l \in\{$Brazoria, Fort Bend, Galveston, Harris, Jefferson$\}$ \\
$t$ & an indicator of time on the total period $T$. $t \in T$ \\
$\mathcal{V}_i$                & A node set of the $i$-th layer. $\mathcal{V}_i = \{v_{i1},v_{i2},\mathellipsis,v_{ij}\}$ $\forall i \in \{h,s,p\}$ and $j \in \{1,..,|\mathcal{G}_i|\}$ \\
$\mathcal{E}_i$                & $\mathcal{E}_i =\{e_{i1},e_{i2},\mathellipsis,e_{ij}\}$ $ \forall~i\in\{h,s,p\}$ and $j \in \{1,\mathellipsis,|\mathcal{G}_i|\}$                               \\
$\mathcal{E}_{ik}$ & $\mathcal{E}_{ik} = \{e_{ik1},e_{ik2},\mathellipsis,e_{ikj}\}$ $ \forall i,k \in \{h,s,p\}$  and $j \in \{1,\mathellipsis,|\mathcal{G}_i|\}$                                                                            \\
$\mathcal{G}_h, \mathcal{G}_s, \mathcal{G}_p$    & $\mathcal{G}_h=\{\mathcal{V}_h, \mathcal{E}_h\}$                                                                                           \\
$\mathcal{G}$                   & $\mathcal{G}=\{\mathcal{G}_h,\mathcal{G}_s,\mathcal{G}_p\}$       \\
$\delta_{ij}$ & A threshold of the $i$-th layer and the $j$-th human agent, $\forall i \in \{h,s,p\}$ and $\forall j \in \{1,\mathellipsis,|\mathcal{G}_i|\}$. \\
$q_i(t)$ & A functionality of the $i$-th layer at time $t$. \\
\hline
\end{tabular}
\label{table_notations}
\end{table*}

\subsection{Definitions of MN and agents}
This section rewrites the definitions of the MN and agents coming from the previous studies to clarify the connections between interactions of this study~\cite{xue2023supporting}.
The ABM-MN simulation framework consists of the MN ($\mathcal{G}=\{\mathcal{G}_h, \mathcal{G}_s, \mathcal{G}_p\}$), agents ($\mathcal{V}=\{\mathcal{V}_{h},\mathcal{V}_{s},\mathcal{V}_{p}\}$), and agent interactions ($\mathcal{E}=\{\mathcal{E}_{h},\mathcal{E}_{s},\mathcal{E}_{p},\mathcal{E}_{hs},\mathcal{E}_{hp}, \mathcal{E}_{sp}\}$). This study defines a household as a human agent in $\mathcal{G}_h$ ($v_{hj}$), a POI as a POI agent in $\mathcal{G}_s$ ($v_{sj}$), a total physical infrastructure facility (water/sewer systems in this study) as a physical infrastructure agent in $\mathcal{G}_p$ ($v_{pj}$). Each node ($v_{hj}, v_{sj}, v_{pj}$) has a corresponding value ($q_{hj}, q_{sj}, q_{pj}$) to represent the recovery state of each node. Note that $q_{hj}$ and $q_{sj}$ have a value ranging from 0 to 1. 

There are six possible types of agent interactions, classified with intra-layer edges and inter-layer edges. Intra-layer edges represent the interactions between agents within one layer ($\mathcal{E}_h,\mathcal{E}_s,\mathcal{E}_p$). Inter-layer edges depict the interactions between agents across the layers ($\mathcal{E}_{hs},\mathcal{E}_{hp}, \mathcal{E}_{sp}$). This study generates intra-layer edges by Tobler's First Law of Geography~\cite{xue2023supporting,tobler2004first,tobler1970computer}, and inter-layer edges by the estimation result that households' return decisions are affected by the neighborhoods, physical infrastructures, and social infrastructures. This study assumes that 1.609 kilometers (1 mile) is a distance that can affect the household's return decision. Note that all underlying mechanisms about the intra- and inter-layer edges are described by~\cite{xue2023supporting}.

This study uses the household return decision models to explain all possible agent interactions. The return decision model covers $\mathcal{E}_{h}$, $\mathcal{E}_{hp}$, and $\mathcal{E}_{hs}$ to isolate one interaction from each other interaction. On the other hand, The SD model simulates the dynamics of the three different types of interactions ($\mathcal{E}_{s}$, $\mathcal{E}_{p}$, and $\mathcal{E}_{sp}$). The SD model integrates the three interactions into a system of ordinary differential equations (ODEs) so that the interactions are interchangeable in the SD model. Note that $\mathcal{E}_p$ only has a recursive link since this study only allows the self dynamics of $\mathcal{V}_p$ due to the lack of data. In short, graphical representations of the MN and the corresponding interactions are shown in Figure \ref{fig1}a. All details, including reasoning, basic statistics, and underlying mechanisms about the MN and agents, are in~\cite{xue2023supporting}.

\subsection{The data-driven SD model in the ABM-MN}
This section rewrites a function of the data-driven SD models in ABM-MN coming from the previous study~\cite{xue2023supporting} to briefly restate the SD model mainly used in this study. Recall that the SD model aims to simulate the three different types of agent interactions, ($\mathcal{E}_{s}$, $\mathcal{E}_{p}$, and $\mathcal{E}_{sp}$) exterior to the agent's return decision rule. This paper selects the data-driven SD model to represent the social ($\mathcal{E}_{s}$), physical ($\mathcal{E}_{p}$), and socio-physical edges ($\mathcal{E}_{sp}$). A part of the socio-physical system reacts to the state given by the components of the ABM-MN. In contrast, the remaining part of the socio-physical system is determined by human interactions ($\mathcal{E}_{h}$). Therefore, this study detaches human interactions ($\mathcal{E}_{h}$, $\mathcal{E}_{hp}$, and $\mathcal{E}_{hs}$) from the socio-physical system. From the previous study~\cite{xue2023supporting}, the socio-physical system by the data-driven SD model is expressed as a set of ODEs as follows:

\begin{equation}\label{SD}
    \frac{d\bar{q}_s(t)}{dt} = 0.001\beta_s \bar{N} q_s(t) (1-\frac{\bar{q}_s(t)}{K_s})+0.1\beta_p \bar{q}_p(t)(1- \frac{\bar{q}_p(t)}{K_p}),
\end{equation}
where $\bar{N}$ is the average number of intra-layer nodes in $\mathcal{G}_s$, $\bar{q}_i(t)$ is an average of the $i$-th layer's recovery state. The value of $\bar{N}$ is as follows: 139.1 (Harris County), 107.7 (Fort Bend County), 79.9 (Brazoria County), 78.5 (Galveston County), and 70.2 (Jefferson County), respectively~\cite{xue2023supporting}. The main interactions from the social nodes to the physical nodes are represented by the logistic relationships~\cite{xue2023supporting,park2022post}. This study estimates the two sets of ODEs to identify the urban-rural differences (See~\cite{xue2023supporting} for the underlying mechanisms and reasoning for the data-driven SD model). The previous study utilized the parameters estimation results from the survey data and the POI data~\cite{xue2023supporting}. Table \ref{table_SDparams} shows the estimated results from the previous paper~\cite{xue2023supporting}.

\begin{table*}[h]
\centering
\footnotesize
\caption{Estimated parameters of SD. This table is cited from ~\cite{xue2023supporting}.}
\begin{tabular}{c|ccccc}
\hline
Parameters       & $\beta_s$ & $K_s$  & $\beta_p$ & $K_p$  \\
\hline
Values (Harris County) & 0.026 & 0.671 & 1.432 & 0.901 \\
Values (Other counties) & 0.093   & 0.736 & 1.114   & 0.935 \\
\hline
\end{tabular}
\label{table_SDparams}
\end{table*}

\subsection{The agent's return decision models}
The agent's return decision model determines the households' return behavior in the PDR process. Recall that the agent's return decision model covers human interactions ($\mathcal{E}_{h}$, $\mathcal{E}_{hp}$, and $\mathcal{E}_{hs}$). This study allows only two options for household returns, not returned (0) or returned (1). There are two types of agent's return decision models: (1) the binary logit model (BLM) and (2) the threshold model. 

\subsubsection{The binary logit model}
The BLM explains a logistic relationship between the independent variables (states of layers and the agent's attributes) and the dependent variable (the return decision). The BLM is widely used to explain human interactions, including the agent's behavioral rules. This paper estimates the BLM by the Post-hurricane household survey. This paper utilizes (1) data filtering, (2) missing data imputation, and (3) the model selection process by the R-value. Table \ref{table_estimation} shows the estimation results. The other details on the survey and corresponding model estimation results are in the previous study~\cite{xue2023supporting}. 

\begin{table*}[h]
\centering
\footnotesize
\caption{Return decision models for the toy network (${}^{**}: <0.001$, and ${}^{*}: <0.01$). This table is cited from ~\cite{xue2023supporting}.}
\begin{tabular}{cccc|cccc}
\hline
\multicolumn{4}{c|}{Urban counties}  & \multicolumn{4}{c}{Rural counties}\\
\hline
Variable                      & Coefficient                                     & Standard Error                                  & p-value                           & Variable                      & Coefficient                                     & Standard Error                                  & p-value                               \\ 
\hline
Intercept                     & $-1.904$                                          & $0.419$                                           & $<0.001^{**}$ & Intercept                     & $-2.379$                                          & $0.682$                                           & $<0.001^{**}$ \\
$q_{house}$                     & $1.520$ & $0.508$ & $0.003^{**}$             & $q_{income}$                     & $2.26\times10^{-5}$ & $8.82\times10^{-6}$ & $0.010^{*}$             \\
$q_{h}$                      & $1.638$                                           & $0.776$                                           & $<0.035^{*}$             & $q_{h}$                      & $3.298$                                           & $1.426$                                           & $<0.021^{*}$             \\
$q_{s}$                     & $-1.756$                                          & $0.565$                                            & $<0.002^{**}$            & $q_{s}$                     & $-4.845$                                          & $1.85$                                            & $<0.009^{**}$\\
$q_{p}$                   & $1.171$                                           & $0.490$                                           & $<0.017^{*}$            & $q_{p}$                   & $1.675$                                           & $0.567$                                           & $<0.003^{**}$            \\
\hline
$n$                             & $99$                                              &                                                 &                                      & $n$                             & $71$                                              &                                                 &                                      \\

$R^2$          & $0.416$                                           &                                                 &                                      & $R^2$          & $0.398$                                           &                                                 &                                      \\
Adj. $R^2$ & $0.342$                                           &                                                 &                                     & Adj. $R^2$ & $0.297$                                           &                                                 &                                     
\\
\hline
\end{tabular}
\label{table_estimation}
\end{table*}

\subsubsection{The threshold model}
The threshold model is a model that can be determined by the threshold values. Instead of using the utility theory~\cite{ben1985discrete}, the threshold model does not permit the commensurability of attributes, isolating each attribute from the other attributes. The generic mechanism of the threshold model is that the three layers ($\mathcal{G}_h, \mathcal{G}_s, \mathcal{G}_p$) have recognized values ($\Tilde{q}_{hj}$, $\Tilde{q}_{sj}$, $\Tilde{q}_{pj}$) to identify the recovery state of the nodes of three layers from the human agents ($v_{hj}, v_{sj}, v_{pj}$), respectively. When $q_{hj}, q_{sj}, q_{pj}$ are higher than the corresponding threshold values ($\delta_{hj}, \delta_{sj}, \delta_{pj}$) respectively, $q_{hj}, q_{sj}, q_{pj}$ set to be 1. Otherwise,  $q_{hj}, q_{sj}, q_{pj}$ set to be 0. 

The underlying mechanism of the threshold model is based on the way of recognizing "human differences in perception"~\cite{ditterich2010comparison}, distinguishing between the human's recognized values ($\Tilde{q}_{hj}$, $\Tilde{q}_{sj}$, $\Tilde{q}_{pj}$) and the actual recovery states of the three layers ($q_{hj}$, $q_{sj}$, $q_{pj}$). $q_{hj}$, $q_{sj}$, and $q_{pj}$ are continuously ranged between 0 and 1. On the other hand, human's recognized values,  $\Tilde{q}_{hj}$, $\Tilde{q}_{sj}$, and $\Tilde{q}_{pj}$ are binary. This paper admits skipping the usage of $\Tilde{\cdot}$ for simplicity.

There are several versions of the threshold model, homogeneous, heterogeneous, stochastic, and time-varying threshold models~\cite{macy2020threshold}. This paper specifies four different types of threshold models, (1) a universally homogeneous threshold model, (2) a universally heterogeneous threshold model, (3) an individually heterogeneous threshold model, and (4) a universally time-varying threshold model to reveal the impact of the threshold model on the ABM-MN. First, the universally homogeneous threshold model represents that all human agents use the same threshold to determine the state of each layer. The universally homogeneous threshold mathematically implies as follows.

\begin{equation}
    \delta_{ij}=\delta_{i}=\delta.
\end{equation}

Second, the universally heterogeneous threshold model moderates the strict condition that all layers have the same threshold model. Contrary to the universally homogeneous threshold model, the universally heterogeneous threshold model allows the different thresholds by different layers. This relaxation explains the heterogeneity between the MNs. The universally heterogeneous threshold model is mathematically expressed as follows.

\begin{equation}
    \delta_{h} \neq \delta_{s} \neq \delta_{p}.
\end{equation}

Third, the heterogeneous threshold model relaxes an assumption that all human agents share the same thresholds for each layer. This paper assumes that the $j$-th heterogeneous threshold model is normally distributed within each layer's mean of recovery levels from the Harvey survey data.
The heterogeneous threshold model is mathematically described as follows.

\begin{equation}
    \delta_{i1}\neq \delta_{i2} \neq \mathellipsis \neq \delta_{ij}.  
\end{equation}

\begin{equation}
    \delta_{hj}\sim N(0.85,0.2), \delta_{pj}\sim N(0.93,0.2), \delta_{sj}\sim N(0.93,0.2)
\end{equation}

The heterogeneous threshold model is theoretically applicable to the real world since the survey data shows the heterogeneous conditions of the attributes for the return decision-see Section \ref{sensitivity} on sensitivity analysis on the population sizes of ABMs. 

The last threshold model is the time-varying threshold model. Time-varying threshold model changes the threshold over time. The model is suitable for expressing the threshold change over time by the agent. 
This study mathematically expresses the time-varying threshold model as follows.

\begin{equation}
    \delta_{ij1} \neq \delta_{ij2} \neq \mathellipsis \neq \delta_{ijt}.
\end{equation}

\subsection{Updating rules of ABM-MN}
ABM-MN updates the agent's elements at every time step (days). At every time step, ABM-MN sequentially updates $q_p$, $q_s$, and $q_h$. $\mathcal{G}_p$ renews the value from the external survey data~\cite{xue2023supporting}. On the other hand, the SD models govern the interactions on $\mathcal{G}_s$ over time. Eq. \ref{SD} has been revised to allow the heterogeneous interactions of $q_{sj}$. Finally, ABM-MN updates $\mathcal{G}_h$ based on the updated result of $q_p$, $q_s$ for the agent's return decision model. Note that this study does not allow the second evacuation after the initial return for the simplicity of the model. This study skips the overall details and assumptions in~\cite{xue2023supporting} since the other details are not closely related to this study.

\section{Experimental setup of ABM-MN}
This study utilizes a two-factor experimental setup to directly quantify the effects of the types of the agent's return decision models and the population size since this study aims to reveal the impact of the types of return decision rules and the population size in the PDR process. The first factor is the types of return decision models. 
This study selects five types of return decision models, (1) the BLM, (2) the universally homogeneous threshold model, (3) the universally heterogeneous threshold model, (4) the individually heterogeneous threshold model, and (5) the universally time-varying threshold model. This study spreads the different threshold values from 0.6 to 0.9 by 0.1 to simulate the various threshold values of each layer. This study iterates the simulation by five other random seeds.

The details of the models are described as follows.
\begin{enumerate}
    \item The universally homogeneous threshold model: $\delta=0.6, 0.7, 0.8$ or $0.9$
    \item The universally heterogeneous threshold model: $\delta_h=0.85$ and $\delta_p=\delta_s=0.93$. 
    \item The individually heterogeneous threshold model: $\delta_{hj}\sim N(0.85,0.2), \delta_{pj}\sim N(0.93,0.2)$, $\delta_{sj}\sim N(0.93,0.2)$.
    \item The universally time-varying threshold model:
    Please refer to Table \ref{table_time-varying}.
\end{enumerate}

\begin{table*}[h]
\centering
\footnotesize
\begin{tabular}{c|ccccc}
\hline
Description & 0 & 3 days & 1 week & 1 month \\
\hline
$q_p$     & 0.57 & 0.78  & 0.97  & 0.89   \\
$q_s$     & 0.68 & 0.98  & 0.97  & 0.89   \\
$q_h$     & 0.70 & 0.91  & 0.94  & 0.81    \\
\hline
\end{tabular}
\caption{\label{table_time-varying} Time-varying thresholds over time.}
\end{table*}

The second factor is the population size. This study analyzes the population ranging from 100 to 5000 by 100. Note that this study excludes a population size of less than 100 since the population does not offer a stable outcome due to the lack of human agents.

Throughout this experimental setup, this study quantifies the impact of types of the agent's return decision models and the population size. 

\subsection{Measuring the CTs by the sensitivity analysis}
The scenario setup in the previous section is one of the bases for finding the conditions for the CTs in the PDR process. Recall that CTs are phenomena of abrupt changes from one state to another by the small perturbation of the elements. Kuehn (2011)~\cite{kuehn2011mathematical} summarizes the common attributes of the CTs on the dynamical systems from Scheffer et al. (2009)~\cite{scheffer2009early} and Scheffer (2020)~\cite{scheffer2020critical},
\begin{enumerate}
    \item "A sudden qualitative change emerges from the system"~\cite{kuehn2011mathematical},
    \item "The change is faster than the usual dynamics"~\cite{kuehn2011mathematical},
    \item "There is a special threshold near the CTs"~\cite{kuehn2011mathematical}, and
    \item "The new value of the system is different from the previous state."~\cite{kuehn2011mathematical}.
\end{enumerate}

Also, Kuehn (2011) notes the assumption of the generic indicators of the CTs~\cite{kuehn2011mathematical} as follows.

\begin{enumerate}
    \item "There is small noise in the major deterministic component of the system," and
    \item "The system slowly returns to the original state from perturbations, called a critical slowdown."
\end{enumerate}

This study intends to use qualitative and quantitative ways to identify CTs. The qualitative way is to check whether the area or point satisfies the CTs' four attributes and two generic indicators. 
The other way to quantitatively determine and measure the CTs on the sensitivity analysis is to select a residual measure to check the difference. Previous studies on the MN organized the mathematical condition for the CTs by the difference of the 1-dimensional system~\cite{boccaletti2014structure}. This paper interprets the difference of the 1-dimensional system as the average of residuals (AR) for the CTs by the sensitivity analysis. For example, this study compares the current curve to the average curve coming from all different population sizes of each threshold. The AR represents the average differences from the selected curve to the average curve. This study only visualizes the averaged values of $q_{hj}$ to equally measure and identify the AR irrespective of the model types. The AR can be described as follows. 

\begin{equation}
    AR_{selected}^{types}=\frac{1}{|T|}\sum_t(q_{selected}^{types}(t)-q_{mean}(t)),
\end{equation}
where $q_{selected}^{types}(t)$ is the human value that represents the simulation selected by different model types and $q_{mean}(t)$ is the human value that represents the model type averaging five different simulation results.  
\section{Results}
\begin{figure}[ht]
    \centering
    \includegraphics[scale=0.46]{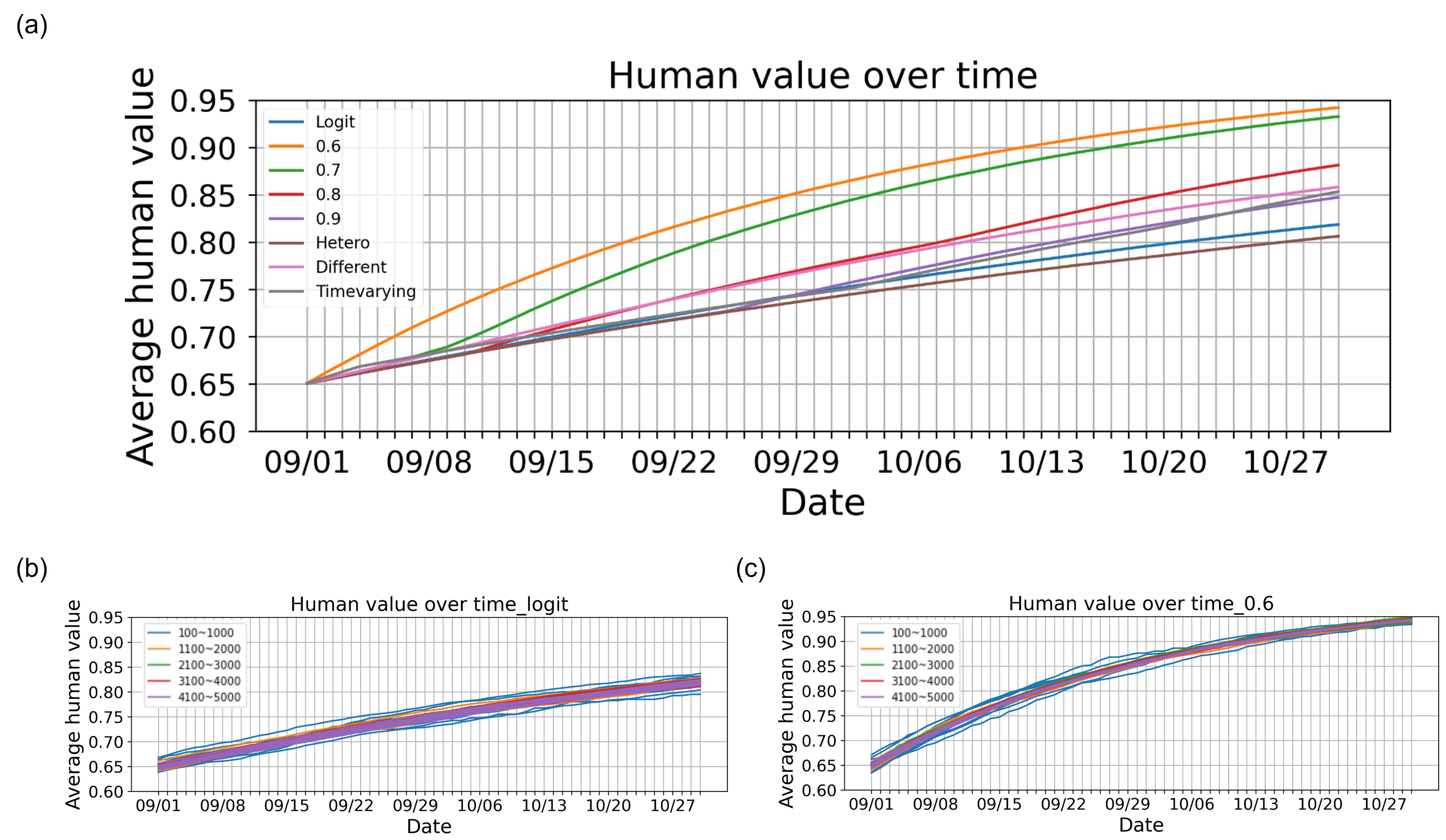}
    \caption{Simulation results over two months by different types of human interactions and population sizes after September 1, 2017. \textbf{(a)} Simulation results aggregated by model types. The blue line shows the logit model results. The orange, green, red, and purple lines show a value of 0.6, 0.7, 0.8, and 0.9 of the universally homogeneous threshold. The dark purple line delineates a universally heterogeneous threshold model (hetero). The pink line represents an individually heterogeneous threshold model (different). The grey line visualizes a universally time-varying threshold model (time-varying). \textbf{(b)} and \textbf{(c)} Examples of simulation results by the BLM and the homogeneous threshold model (0.6). The blue line represents the human values when the population ranges from 100 to 1000. The other colored lines represent the human values ranging from the first value to the last value.}
    \label{fig2}
\end{figure}

Figure \ref{fig2} represents the simulation results over two months by different types of interactions and population sizes. Figure \ref{fig2}a shows average human values aggregated by the eight types of agent's return decision models. Recall that human value ($q_{hj}$) represents the human recovery state of agent $j$ ranging from 0 (evacuated) and 1 (returned). Each curve in Figure \ref{fig2}a represents the aggregation of the curves with a combination of 5 different random seeds and 50 different population sizes. The universally homogeneous threshold models (0.6 and 0.7) are higher than the BLM model curve, which implies that the low universally homogeneous threshold model does not clearly explain the BLM results.

The BLM model simulation results are in Figure \ref{fig2}b, and the universally homogeneous threshold (0.6) model results are in Figure \ref{fig2}c. Figures \ref{fig2}b and \ref{fig2}c show the variability of curves with 100 to 1000 populations, while the other four types of curves (1,100 to 2,000, 2,100 to 3,000, 3,100 to 4,000, and 4,100 to 5,000 populations) are gathered. It implies that curve variability happens at the curves with less than 1,000 populations.

\subsection{Sensitivity analysis on the population sizes of ABMs}\label{sensitivity}
\begin{figure}[h]
    \centering
    \includegraphics[scale=0.5]{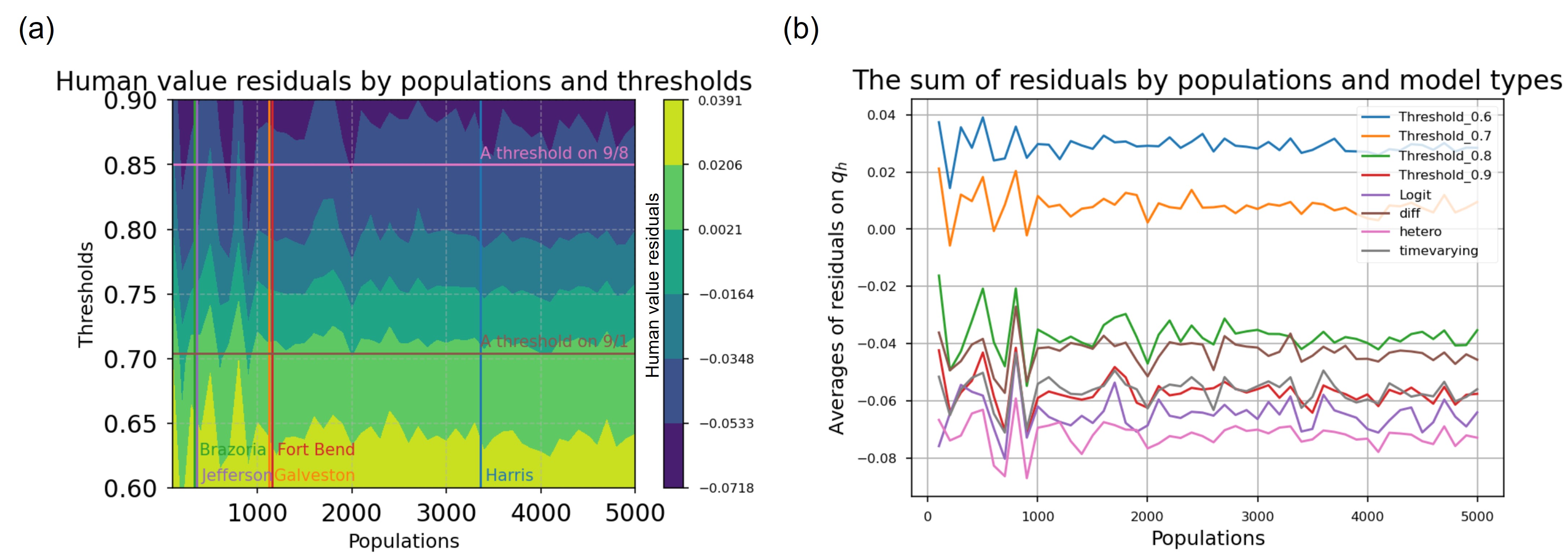}
    \caption{Sensitivity analysis of the ABM-MN. \textbf{(a)} The average of human value residuals by populations and thresholds.  \textbf{(b)} The sum of residuals by different types of interactions.
}
    \label{fig3}
\end{figure}

Figure \ref{fig3}a shows the sensitivity analysis of the ABM-MN by the population sizes and model types. This plot uses the AR to identify the differences by the result of the homogeneous threshold model. The vertical solid line represents the population density of the selected counties, while the horizontal solid line means the threshold value on September 1st and September 8th, 2017. The bright area of this plot implies a higher value of AR than that of the other areas. 

The overall patterns are horizontally placed, meaning that threshold change affects the significant value change, while population change affects the minor value change. However, when you check the population of less than 1,000, they show an abrupt change in the given conditions. It implies that the system frequently changes the AR. Populations over 1,000 are relatively stable compared to those of less than 1,000. It leads to the idea that less than 1,000 population results should be somewhat vulnerable to abrupt change, yielding the CTs. Also, high thresholds make a relatively significant difference compared to thresholds over 0.75 when we divide the area into minor and significant differences by the median of the AR. Therefore, the possible CT condition ranges from the thresholds between 0.6 to 0.75 and a population of less than 1,000.

Figure \ref{fig3}b compares the SR by the other agent's return decision model types. Compared to the ARs of two universally homogeneous threshold model curves (Threshold\_0.6 and Threshold\_0.7), the other five curves are within the negative range between -0.08 and -0.02, implying the similarity among the five curves. Parts of each curve with less than 1,000 populations have been perturbed, supporting the idea that less than 1,000 populations are vulnerable to the CTs supported by the generic indicators. Additionally, it also satisfies the attributes of the CTs. Therefore, we should identify the CTs' region. The result is aligned with Figure \ref{fig2}.

\subsection{Case studies: The five selected counties}
\begin{figure}[h]
    \centering
    \includegraphics[scale=0.55]{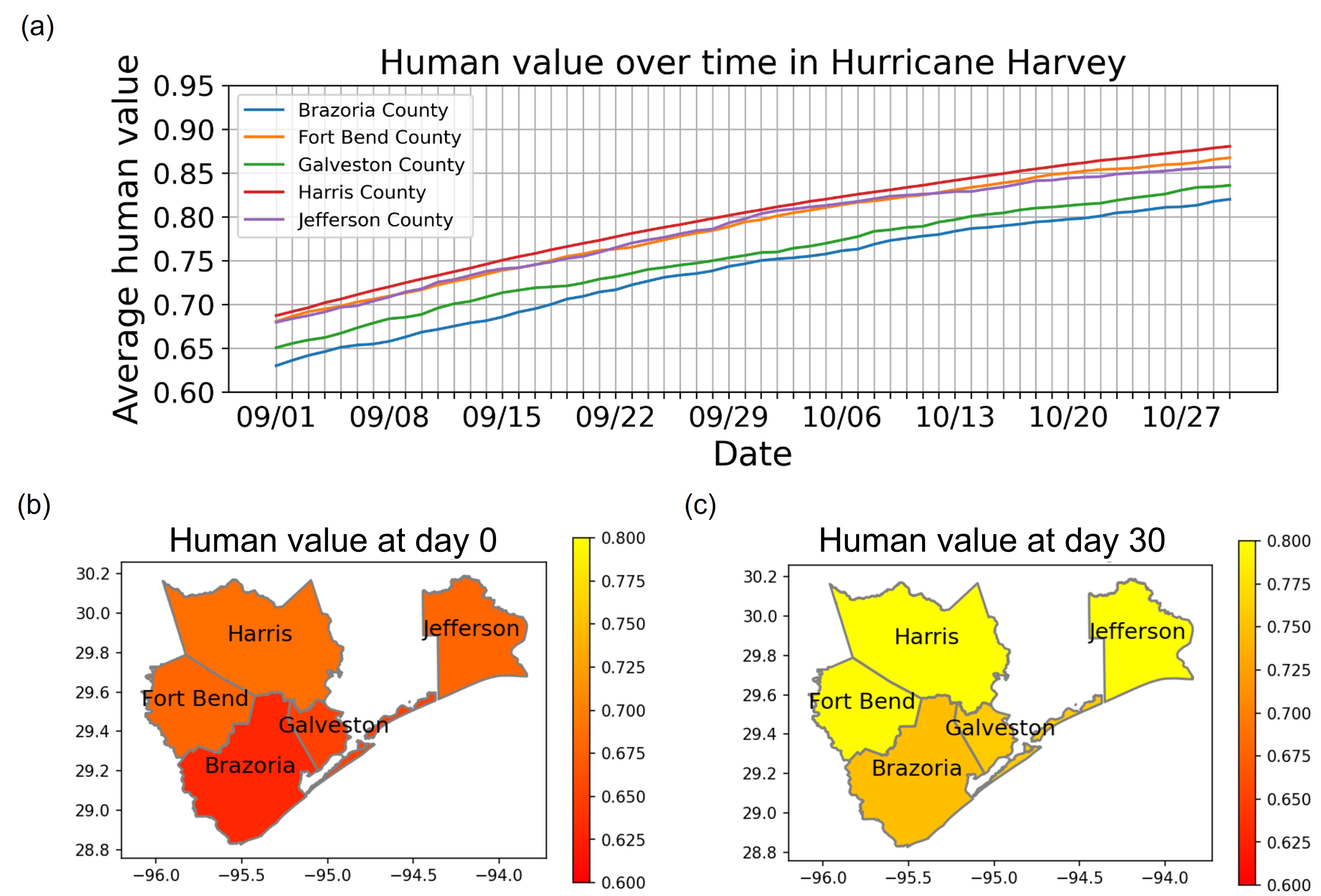}
    \caption{Real-world simulation results. \textbf{(a)} Average human values over two months by the model types in five selected counties in Hurricane Harvey. \textbf{(b)} Visualization of human value on five counties at day 0. \textbf{(c)} Visualization of human value on five counties on the day 30.}
    \label{fig4}
\end{figure}

\begin{figure}[p]
    \centering
    \includegraphics[scale=0.55]{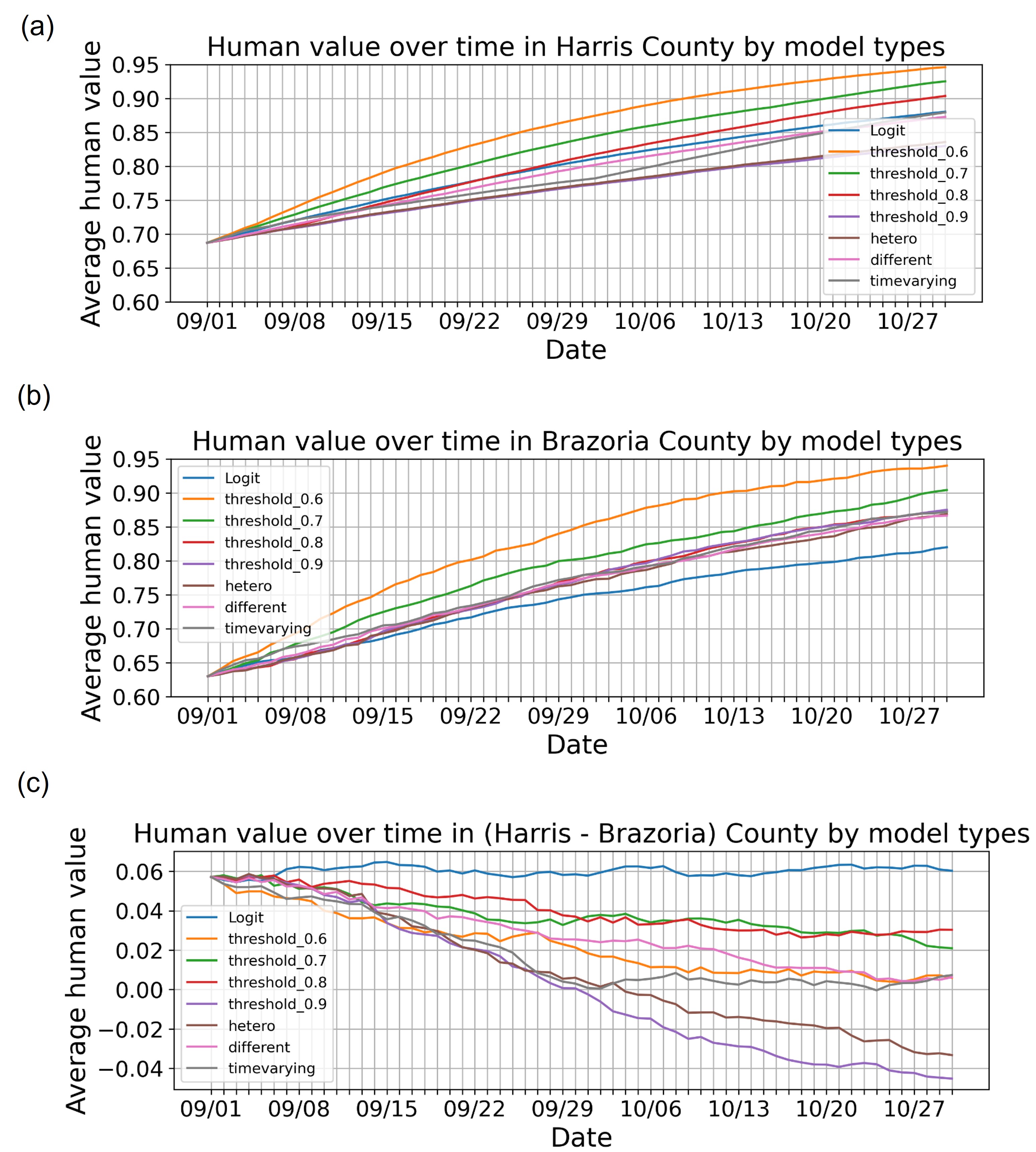}
    \caption{Urban-Rural differences in the simulation results. \textbf{(a)} Human value over time in Harris County by agent's return decision model types. \textbf{(b)} Human value over time in Brazoria County by agent's return decision model types. \textbf{(c)} The human value difference between Harris County and Brazoria County.}
    \label{fig5}
\end{figure}

This study applies ABM-MN to the five selected counties. Figures \ref{fig4} and \ref{fig5} represent the ABM-MN simulation results on the selected counties. 
Figure \ref{fig4}a shows average human values by different counties. This plot uses the averages of the eight agent's return decision model types. The overall curves are relatively stable compared to the toy network results, but the general human values are lower than the toy network results. Brazoria County and Galveston County are somewhat lower than those of the other three counties. We can also check the low average human value on September 1st and October 1st from Figures \ref{fig4}b and \ref{fig4}c. 

Figures \ref{fig5}a and \ref{fig5}b represent the averaged human values by types of agent's return decision models on Harris and Brazoria Counties. We found that changes in the agent's return decision models dramatically affect the average human value. This plot shows the 8.9\% of changes in the average human value on 10/31 when we choose the value with 0.6 of the universally homogeneous threshold (threshold\_0.6) and the value with the BLM (Logit). The universally heterogeneous threshold model (hetero), an individually heterogeneous threshold model (different), and a universally time-varying threshold model (timevarying) are close to the BLM model. Still, they are relatively flexible in the first few weeks. 

\subsubsection{Identification of urban-rural differences}
Figure \ref{fig5} shows the urban-rural differences by the agent's return decision model types. Figure \ref{fig5}a visualizes the urban county (Harris County) cases, and Figure \ref{fig5}b shows the rural county (Brazoria County) cases. Figure \ref{fig5}c shows the difference between Harris and Brazoria Counties by the agent's return decision model types. The BLM results indicate that the human value averages of the urban county are higher than those of the rural county. On the other hand, the universally homogeneous threshold model with the value of 0.9 (threshold\_0.9) and the universally heterogeneous threshold model (hetero) have higher averages of the human values of the rural county than that of the urban county. The other five model types have similar values on the urban-rural differences. Note that the time-varying threshold model (timevarying) can freely switch the tendency of the curve. 

\section{Discussion}
This study compares the averaged human recovery curve to the target curve by the AR to find the CTs of the PDR process in the sensitivity analysis of the ABM-MN. The toy network and real-world cases are selected to reveal the effect. The main factors of this study are the fifty different population sizes and eight different agent's return decision model types. This study not only qualitatively evaluates the CT area by six points from previous studies on the CTs~\cite{kuehn2011mathematical}, but also quantitatively validates the CT area by the AR.

This paper found three points: (1) the applicability of the threshold model, (2) the importance of the agent's return decision model types and population sizes, and (3) the urban-rural human value differences.

First, this study finds the similarity between the threshold model and the BLM model. These two model types can be interchangeable since they share similar macroscopic patterns in the model types and population sizes in the toy network and the real-world cases. Due to the commensurability of attributes assumed in the BLM model, the threshold model is an alternative to the BLM model, keeping the isolation of non-interchangeable attributes. The variations of the threshold model, such as the individually heterogeneous threshold model and universally time-varying threshold model, can successfully simulate the PDR process with no remarkable signs of differences between them except for the universally low homogeneous threshold model. 

Second, this study suggests the condition to emerge the CTs of the PDR process in the sensitivity analysis. High threshold models and less than 1,000 populations yield the perturbation of the system, satisfying four common attributes of the CTs and two notable indicators of the CTs. High populations do not make the CTs in given conditions caused by Hurricane Harvey. Therefore, policymakers should be cautious about using the lower threshold value models and low population sizes in the ABM-MN.

We can also check the reliability of the results from the limited sample of mobile phone users based on the importance of the population size. If we do not simulate a sufficient number of mobile phone location data compared to the population, ABM-MN results will be in the perimeter of the CTs. Therefore, we should suggest the appropriate synthetic population generation techniques compensating for the tiny proportion of the datasets. Previous ABM studies supported the issues of insufficient population sizes~\cite{srikrishnan2021small}.

\subsection{Identification of urban-rural differences}
Third, this study found the urban-rural differences in different model types. Harris and Brazoria Counties have a significant difference in the overall curves. It may come from the population difference between the two types of counties. The difference also happens in the different model types, although the signs of differences are different on the homogeneous threshold model (threshold\_0.9) and the universally heterogeneous threshold model (hetero). Therefore, policymakers should consider the conditions of the CTs (high threshold models and low population density) when focusing on the PDR process of rural counties. If policymakers overlook the perturbation signs of the rural counties, the system will suffer the CTs of the system, which can be prevented in advance.

\subsection{Limitations}
This study has five limitations. First, this study does not focus on population variability in real-world case studies. Table \ref{table_demographics} shows the insufficient ratio of valid users less than 1\% of the population density. Appropriate techniques of synthetic population generation are necessary to validate the results of ABM-MN. Second, this study does not use the mathematically rigorous definition of CTs and the corresponding conditions. Specific dynamical systems satisfy certain ranges of the CTs, combining the concept of early warning signals. To do this, mathematically rigorous definitions of ABM-MN are required. Mean-field theory-based game theoretic approach across the agents can simplify and boost the ABM-MN sufficiently to analyze the CTs and the corresponding conditions.

Third, this study utilizes the county-level physical infrastructure recovery status data to represent the water system in the PDR process. We can extend the ABM further after obtaining the granular physical infrastructure recovery status. Next, this study ignores the impact of housing damage in ABM-MN due to the lack of data sources. Since housing damage is one of the critical factors for the PDR process, it should be considered to specify the more realistic agent interactions. The last limitation is that AR does not function as the early warning signal of the CTs. Capturing the early warning signals of the CTs improves the system's preparedness for the incident. A new index for catching the early warning signals of the CTs is necessary.

\section{Conclusion}
This study is the following work of the previous study~\cite{xue2023supporting}. Specifically, this paper focuses on the simulation of the PDR process to analyze the impact of the agent's return decision types and population sizes in the ABM-MN. Hurricane Harvey in 2017 and five coastal counties near Houston were selected to estimate the parameters and check the real-world case studies.
This study found that the threshold model can substitute the BLM model free from the commensurability of attributes. Also, this study considers and identifies the conditions of the CTs in the PDR process with high thresholds and less than 1,000 populations in the toy network. This study applies the sensitivity analysis to real-world case studies and checks the urban-rural human value differences by the different agent's return decision model types. 
This study sheds light on introducing the threshold models to the PDR process and finding early warning signals of the CTs in the PDR process.

\section{Authorship Contribution}
\textbf{Sangung Park}: Conceptualization, Methodology, Coding, Figure, and Writing.
\textbf{Jiawei Xue}: Conceptualization, Methodology (ABM-MN), Editing.
\textbf{Satish V. Ukkusuri}: Conceptualization, Methodology, Discussion, Editing.

\section{Data and Code Availability}
The data and code for the ABM are available at \url{https://github.com/Sangung/CRISP_ABM_CT}.

\section{Acknowledgement}
S.P., J.X., and S.V.U. were partly funded by NSF Grant No. 1638311, CRISP Type 2/Collaborative Research: Critical Transitions in the Resilience and Recovery of Interdependent Social and Physical Networks. 

\bibliographystyle{trb}
\bibliography{ref}
\end{document}